\documentclass[aps,pre,onecolumn,notitlepage,groupedaddress]{revtex4-1}
\usepackage{epsfig,graphics,amssymb,amsmath,subeqnarray,color}
\usepackage{siunitx,pifont}
\definecolor{navyblue}{RGB}{0,0,128}
\definecolor{dodgerblue}{RGB}{30,144,255}
\definecolor{darkgrey}{RGB}{169,169,169}
\definecolor{skyblue}{RGB}{135, 206, 235}

\begin{document}

\title{Squirming through shear-thinning fluids}
\author{Charu Datt}
\affiliation{Department of Mechanical Engineering, University of British Columbia,\\
Vancouver BC, V6T 1Z4, Canada}
\author{Lailai Zhu}
\affiliation{Laboratory of Fluid Mechanics and Instabilities, \'Ecole Polytechnique F\'ed\'erale de Lausanne,\\ 
Lausanne, CH-1015, Switzerland}
\author{Gwynn J. Elfring}\email{Electronic mail: gelfring@mech.ubc.ca}
\affiliation{Department of Mechanical Engineering, University of British Columbia,\\
Vancouver BC, V6T 1Z4, Canada}
\author{On Shun Pak}\email{Electronic mail: opak@scu.edu}
\affiliation{Department of Mechanical Engineering, Santa Clara University,\\
Santa Clara, CA, 95053, USA}


\begin{abstract}
Many microorganisms find themselves immersed in fluids displaying non-Newtonian rheological properties such as viscoelasticity and shear-thinning viscosity. The effects of viscoelasticity on swimming at low Reynolds numbers have already received considerable attention, but much less is known about swimming in shear-thinning fluids. A general understanding of the fundamental question of how shear-thinning rheology influences swimming still remains elusive.  To probe this question further, we study a spherical squirmer in a shear-thinning fluid using a combination of asymptotic analysis and numerical simulations. Shear-thinning rheology is found to affect a squirming swimmer in nontrivial and surprising ways; we predict and show instances of both faster and slower swimming depending on the surface actuation of the squirmer. We also illustrate that while a drag and thrust decomposition can provide insights into swimming in Newtonian fluids, extending this intuition to problems in complex media can prove problematic.
\end{abstract}

\maketitle

\section{Introduction}
Self-propulsion at small length scales is widely observed in biology, common examples include spermatozoa reaching the ovum during reproduction, microorganisms escaping predators and microbes foraging for food \citep{Bray2000, fauci2006biofluidmechanics}. While swimming at low Reynolds numbers is well studied for Newtonian fluids \citep{lauga2009hydrodynamics}, an understanding of the effects of complex (non-Newtonian) fluids on locomotion is still developing. Many biological fluids, such as blood or respiratory and cervical mucus, display complex rheological properties including viscoelasticity and shear-thinning viscosity \citep{merrill1969rheology, Mucus_rheology}. A viscoelastic fluid retains a memory of its flow history, whereas the viscosity of a shear-thinning fluid decreases with shear rate. While it is important to elucidate how non-Newtonian fluid rheology influences propulsion at low Reynolds numbers because microorganisms swim through biological fluids possessing these properties, an improved understanding may also guide the design of artificial micro-swimmers \citep{qiu2014swimming} and novel micro-systems \citep{Mathijssen2015} exploiting these nonlinear fluid properties.

Recent research has begun to shed light on the effects of viscoelasticity (see the reviews by \citet{sznitman15} and \citet{Gwynn_book}), but much less is known about swimming in shear-thinning fluids at low Reynolds numbers. \citet{moumita_2013} measured a decreased swimming speed of a waving sheet in a shear-thinning viscoelastic fluid relative to a Newtonian fluid. In contrast, an asymptotic study of a sheet driven by small-amplitude waves showed that the swimming speed of a waving sheet remains unchanged in an inelastic shear-thinning fluid compared to a Newtonian fluid \citep{Rodrigo_lauga}. A recent experiment by \citet{Arratia_JFM} on the locomotion of the nematode \textit{Caenorhabditis elegans} has also suggested that shear-thinning viscosity does not modify the nematode's beating kinematics or swimming speed. In addition, numerical studies \citep{montenegro2012modelling, Montenegro_shearthinning} examined a variety of two-dimensional swimmers and showed that faster or slower swimming in shear-thinning fluids can occur depending on the class of swimmer and its swimming gait. The results were understood in terms of the fluid viscosity distribution surrounding the thrust and drag elements of the swimmer. By estimating separately the propulsive thrust and drag force on the swimmer, \citet{qiu2014swimming} obtained a scaling relation predicting the swimming velocity of a single-hinge swimmer (a micro-scallop), which is enabled to move at low Reynolds numbers by shear-thinning rheology.

The question that emerges from recent literature is when (and why) does a swimmer go faster or slower in a shear-thinning fluid \citep{gooeyness}. To address this question we study a canonical idealized model swimmer, the squirmer, in a shear-thinning fluid described by the Carreau-Yasuda model using a combination of asymptotic analysis and numerical simulations.We predict and show instances of both faster and slower swimming depending on the surface actuation of the squirmer. We also explore separately the effects of shear-thinning on the propulsive thrust generated by the squirmer and the drag force it experiences, and demonstrate that extending these findings to swimming in non-Newtonian fluids can prove problematic.

\section{Theoretical framework}
The hydrodynamics of spherical bodies propelling themselves with surface distortions, otherwise known as squirmers, was first studied by \citet{lighthill1951} and \citet{Blake1971}. We follow this approach and model a squirmer with prescribed time-independent tangential surface distortions. The resulting slip velocity around the squirmer is decomposed into a series of Legendre polynomials of the form $u_{\theta}(r = a,\theta) = \sum_{l=1} ^{\infty} B_l V_l \left ( \theta \right)$, where $V_l(\theta)=-(2/l (l + 1)) P_l^1 \left( \cos{\theta} \right)$ with 
$P_l^1$ being the associated Legendre function of the first kind and $\theta$ the polar angle measured with the axis of symmetry.

The coefficients $B_l$ are related to Stokes flow singularity solutions. In a Newtonian fluid, the $B_1$ mode (a source dipole) is the only mode contributing to the swimming velocity, and the $B_2$ mode (the stresslet) is the slowest decaying spatial mode and thus dominates the far field velocity generated by squirmers. Therefore, often only the first two modes, $B_1$ and $B_2$, of the expansion are considered \citep{Ishikawa2006, Downton2009,  Yazdi2015}. The ratio of the two modes, $\alpha = B_2 / B_1$, characterises the type of swimmer in a Newtonian fluid: $\alpha>0$ describes a puller, which generates impetus from its front end (\textit{e.g.~}the alga \textit{Chlamydomonas}), whereas $\alpha<0$ represents a pusher, which generates propulsion from its rear part (\textit{e.g.~}the bacterium \textit{Escherichia coli}), and the $\alpha = 0$ case corresponds to a neutral squirmer which induces a potential velocity field. In a Newtonian fluid, the swimming speed of a squirmer $U_N=2 B_1/3$ \citep{lighthill1951, Blake1971}, which is independent of the fluid viscosity because drag and thrust change equally with viscosity. Any modes other than $B_1$ only modify the surrounding flow structure but do not contribute to the swimming speed of a squirmer. This simple picture, however, does not apply to squirming in a shear-thinning fluid, as we discuss later, where all modes can potentially contribute to the swimming velocity, and adding any other modes to $B_1$ will nontrivially affect the locomotion of the squirmer.

\subsection{Shear-thinning rheology: the Carreau-Yasuda model}\label{sec:CY}
Shear-thinning fluids experience a loss in apparent viscosity with applied strain rates, a property that results from changes in the fluid microstructure. As the rate of strain exceeds the rate of structural relaxation, one observes microstructural ordering in the fluid \citep{brown2011through}. Here, we capture the change in apparent viscosity due to this ordering using the Carreau-Yasuda model for generalised Newtonian fluids \citep{bird1987dynamics}. The variation of viscosity with applied strain rate is given by
\begin{align}
\eta = \eta_{\infty} + \left( \eta_0 - \eta_{\infty}\right) \left[ 1+ \lambda_t ^2 \vert \dot{\gamma} \vert^2\right]^{\frac{n-1}{2}},
\label{carreau_equation}
\end{align}
where $\eta_0$ and $\eta_{\infty}$ are the zero- and infinite-shear rate viscosities respectively. The power law index $n$ characterises the degree of shear-thinning ($n < 1$) and the relaxation time $\lambda_t$ sets the crossover strain rate at which non-Newtonian behaviour starts becoming significant. The magnitude of the strain rate tensor is given by $\lvert\dot{\gamma}\rvert = \left( \Pi /2 \right)^{1/2} $, where $\Pi = \dot{\gamma}_{ij} \dot{\gamma}_{ij}$ is the second-invariant of the tensor. As an example, measured values for human cervical mucus can be fitted by the Carreau-Yasuda model with values $\eta_0  = 145.7 $ \si{\pascal\second}, $\eta_{\infty} = 0 $ \si{\pascal \second}, $\lambda_t = 631.04 $ \si{\second}, $n = 0.27$ \citep{Rheology_mucus, Rodrigo_lauga}.
 
We non-dimensionalise the flow quantities taking the first mode, $B_1$, of the surface actuation as the scale for velocity and radius, $a$, of the squirmer as the characteristic length scale.  The strain rates are scaled with $\omega =   B_1/ a $  and the stresses by $ \eta_0 \omega $, such that  the constitutive equation takes the dimensionless form 
\begin{align}
\boldsymbol{\tau}^* = \left\{ \beta + \left( 1- \beta\right) \left[ 1+ Cu^2 \vert \dot{\gamma}^* \vert ^2 \right]^{\frac{n-1}{2}}\right\} \boldsymbol{\dot{\gamma}}^*,
\label{consti_eq}
\end{align}
where $\boldsymbol{\tau}$ is the deviatoric stress tensor, and dimensionless quantities are denoted by stars (*). The Carreau number $Cu = \omega \lambda_t$ is the ratio of the characteristic strain rate, defined by the surface actuation $\omega$, to the crossover strain rate, defined by the fluid relaxation $1/\lambda_t$. The viscosity ratio is given by $\beta= \eta_{\infty} / \eta_0 \in [0,1]$. 

It is evident from \eqref{consti_eq} that when the actuation rate $\omega$ is much smaller or much larger than the fluid relaxation rate $1/\lambda_t$, \textit{i.e.} when $Cu \rightarrow 0$ or $Cu \rightarrow \infty$, the shear-thinning fluid reduces to a Newtonian fluid of constant viscosity $\eta_0$ (dimensionless viscosity 1) or $\eta_\infty$ (dimensionless viscosity $\beta$) respectively. Recalling that for a given surface actuation the swimming speed of a squirmer in the Newtonian regime is independent of the fluid viscosity,  we therefore expect the swimming speed of a squirmer in a shear-thinning fluid to converge to its Newtonian value in the limits $Cu \rightarrow 0 $ or $Cu \rightarrow \infty$. Non-monotonic variation of the swimming speed with $Cu$ is expected for any swimmer with prescribed kinematics and has been observed by \citet{montenegro2012modelling} for some two-dimensional model swimmers. In this study we employ both asymptotic analysis and numerical simulations to investigate these effects of shear-thinning rheology on swimming at low Reynolds numbers.

\subsection{Asymptotic analysis}
The deviatoric stress tensor $\boldsymbol{\tau}^*$ in \eqref{consti_eq} is a non-linear function of the strain rate tensor $\boldsymbol{\dot{\gamma}}^*$. Assuming only a weak nonlinearity, we may uncouple the Newtonian and non-Newtonian contributions writing 
\begin{align}
\boldsymbol{\tau}^* = \boldsymbol{\dot{\gamma}}^* + \varepsilon \mathbf{A}^*, \label{eq:general}
\end{align}
with $\varepsilon \ll 1$ as a dimensionless measure of the deviation from the Newtonian case ($\varepsilon = 0$).

{We observe that in the limits $Cu = 0$ or $\beta = 1$, \eqref{consti_eq} reduces to a Newtonian constitutive equation.} Thus, one may expect weakly nonlinear behaviour when the fluid relaxation rate is much faster than the surface actuation rate ($\varepsilon = Cu^2 \ll 1$), or when the zero-shear-rate viscosity is very close to the infinite-shear-rate viscosity ($\varepsilon = 1-\beta \ll 1$). 
 
Henceforth we shall work in dimensionless quantities and therefore drop the stars (*) for convenience.

\subsubsection{Expansion in Carreau number} \label{sec:Cu}
Expanding all fields in regular perturbation series in $\varepsilon = Cu^2$, we obtain, order by order, the constitutive equations
\begin{align}
\boldsymbol{\tau}_0 &= \boldsymbol{\dot{\gamma}}_0, \notag \\
\boldsymbol{\tau}_1 &= \boldsymbol{\dot{\gamma}}_1 + \frac{\left(n-1\right)}{2} \left ( 1- \beta \right) \vert \dot{\gamma_0}\vert^2 \boldsymbol{\dot{\gamma}}_0,
\end{align}
hence $\mathbf{A} = \frac{\left(n-1\right)}{2} \left(1 - \beta \right) \vert \dot{\gamma_0}\vert^2 \boldsymbol{\dot{\gamma}}_0$ to leading order in \eqref{eq:general}. Note that the first correction to the Newtonian behaviour is linear in $n$, which points to a linear dependence of the swimming speed in $n$ upon using \eqref{final_U_eq}, elucidating the trend suggested by the two-dimensional numerical findings in \citet{Montenegro_shearthinning}. We also remark that this expansion is valid only when $Cu^2 \vert \dot{\gamma}\vert^2$ is $o(1)$ and is therefore not uniformly valid across all values of strain rates.

\subsubsection{Expansion in viscosity ratio}\label{viscosity_expansion}
Expanding in perturbation series with $\varepsilon = 1 - \beta$ gives us, order by order, the constitutive equations, 
\begin{align}
\boldsymbol{\tau}_0 &= \boldsymbol{\dot{\gamma}}_0,\\
\boldsymbol{\tau}_1 &= \boldsymbol{\dot{\gamma}}_1 + \left\{ -1 + \left( 1+ Cu^2 \vert \dot{\gamma}_0\vert^2\right)^{\frac{n-1}{2}}\right\} \boldsymbol{\dot{\gamma}}_0,
\label{beta_expansion} 
\end{align}
where in this limit $\mathbf{A} = \left\{ -1 + \left( 1+ Cu^2 \vert \dot{\gamma}_0\vert^2 \right)^{\frac{n-1}{2}}\right\} \boldsymbol{\dot{\gamma}}_0$ to leading order in \eqref{eq:general}. Note that this asymptotic expansion is uniformly valid for all strain rates or Carreau numbers, which permits a full-range study of the non-monotonic swimming behaviour.

\subsection{The reciprocal theorem}\label{reciprocal_theorem}
\citet{stone_reciprocal} demonstrated the use of the Lorenz reciprocal theorem in low-Reynolds-number hydrodynamics \citep{happel2012low} to obtain the swimming velocity of a squirmer for a given prescribed surface actuation $\mathbf{u}^S$ without calculation of the unknown flow field, provided one can solve the resistance/mobility problem for the swimmer shape (with surface $S$). \citet{High_Deborah, lauga2014locomotion} then developed integral theorems extending this method for use with complex fluids. We use these methods in the subsequent calculations to obtain the swimming velocity of a squirmer in a shear-thinning fluid, which closely follow the formulation in \citet{Gwynn_book}. 
\par 
We represent the velocity field and the associated total stress tensor for a force- and torque-free swimmer with $\textbf{u}$ and $\boldsymbol{\sigma}$ respectively. We consider the corresponding resistance problem in a Newtonian fluid to simplify the calculation of the swimming velocity.  The resistance problem (denoted with a `hat') involves the rigid-body motion with translational velocity $\hat{\mathbf{U}}$ and rotational velocity $\hat{\boldsymbol{\Omega}}$,
and the corresponding velocity field and associated stress tensor are represented by  $\hat{\mathbf{u}}$ and $\hat{\boldsymbol{\sigma}}$ respectively. Due to the linearity of the Stokes equation, we may write $ \hat{\textbf{u}} = \hat{\mathcal{L}} \cdot \hat{\mathcal{U}}$, $\hat{\boldsymbol{\sigma}} = \hat{\mathcal{T}}\cdot \hat{\mathcal{U}}$ and $\hat{\mathcal{F}} = -\hat{\mathcal{R}}\cdot \hat{\mathcal{U}}$. Here, for compactness,  both the translational and rotational components of velocity are contained in $\hat{\mathcal{U}}$ and, similarly, the corresponding matrices contain both the translational and rotational terms. In weakly nonlinear complex fluids, the six-dimensional
translational and rotational velocity of the swimmer $\mathcal{U}$ is then given by
\begin{align}\label{general_swimming}
\mathcal{U} = \hat{\mathcal{R}}^{-1} \cdot \left[ \int_S \mathbf{u}^S \cdot(\mathbf{n}\cdot \hat{\mathcal{T}})\: \mathrm{d}S - \varepsilon\int_V \mathbf{A} : \nabla \hat{\mathcal{L}}\: \mathrm{d}V \right].
\end{align}
The integral over the volume of fluid $V$ external to $S$ in the equation measures the change in swimming dynamics due to the non-Newtonian behaviour of the fluid. For a spherical squirmer with axisymmetrical tangential surface distortions, there is no rotational motion and the translational velocity is given simply by
\begin{equation}
{\textbf{U}} = -\frac{1}{4\pi} \int_S \textbf{u}^S \mathrm{d}S - \frac{\varepsilon}{8 \pi } \int_V \mathbf{A}: \left( 1 + \frac{1}{6} \nabla^2 \right) \nabla \mathbf{G }\:\mathrm{d}V,
\label{final_U_eq}
\end{equation}
where $\mathbf{G} =  \frac{1}{r}\left( \mathbf{I} + \frac{\textbf{r} \textbf{r}}{r^2}\right)$ is the Oseen tensor (or Stokeslet). The first term on the right-hand side is the result of swimming in a Newtonian fluid  \citep{stone_reciprocal}, and the last term in equation contains the weakly nonlinear effect, which can be evaluated analytically in some special cases and can be computed  in general by numerical quadrature readily.

\subsection{Numerical solution}
The numerical simulations of the momentum equations at zero Reynolds number with the Carreau-Yasuda constitutive relation \eqref{carreau_equation} are implemented in the finite element method software COMSOL. We use a square computational domain of size $500a \times 500a$, discretized by about 30000--50000 Taylor-Hood ($P2-P1$) triangular elements. The mesh is refined near the squirmer in order to properly capture the spatial variation of the viscosity. Since slowly decaying flow fields are  expected at low Reynolds numbers, a large domain size is important to guarantee accuracy. The simulations are performed in a reference frame moving with the swimmer and the far-field (inlet) velocity is varied to obtain a computed zero force on the squirmer. In addition to comparing with the asymptotic analysis in this paper, we have validated our implementation against the analytical results of a three-dimensional squirmer in a Newtonian fluid \citep{lighthill1951, Blake1971} and a two-dimensional counterpart in a shear-thinning fluid \citep{Montenegro_shearthinning}.

\section{Results and discussion}

\begin{figure}
\centering
\begin{minipage}{.5\textwidth}
  \centering
  \includegraphics[scale= 0.26, trim = 0 30 50 70, clip]{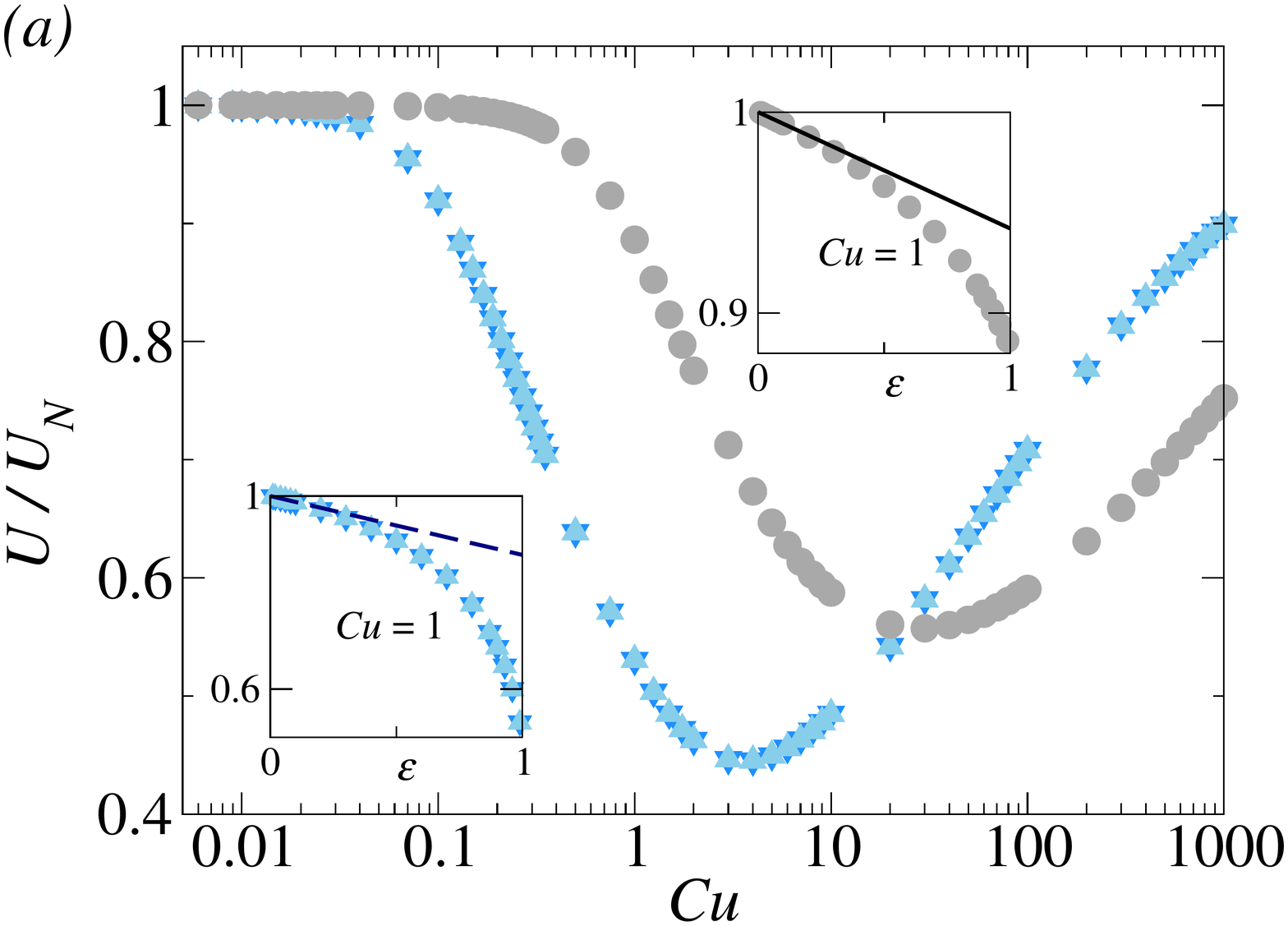}
  \end{minipage}%
\begin{minipage}{.41\textwidth}
  \centering
  \includegraphics[scale= 0.26, trim = 0 30 50 70, clip]{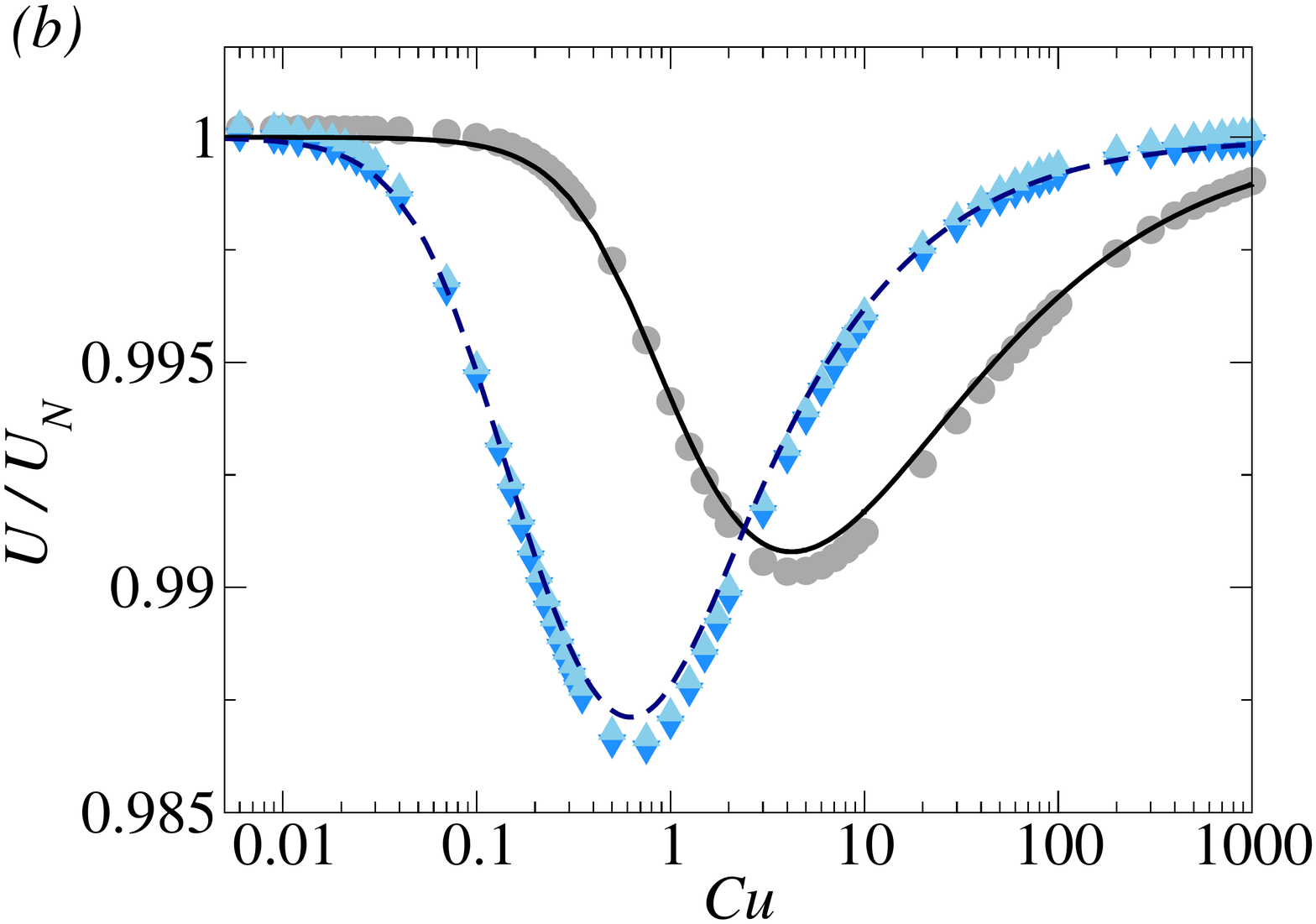}
 \end{minipage}
 \caption{In \textit{(a)} we show results from numerical simulations for a neutral squirmer ($\alpha=0$, {\footnotesize\textcolor{darkgrey}{\ding{108}}}),  puller ($\alpha=5$, {\footnotesize \textcolor{skyblue}{\ding{115}})} and pusher  ($\alpha=-5$, {\footnotesize \textcolor{dodgerblue}{\ding{116}})} for $\varepsilon = 0.99 $ and $n = 0.25$. \textit{(b)} The non-monotonic variation in velocity is well captured by asymptotics for a neutral squirmer (solid line), and a pusher/puller (dashed line)  at $\varepsilon = 0.1$ and $n = 0.25$. 
}
\label{fig1}
\end{figure}

As a first step we investigate the effect of shear-thinning rheology upon swimming speed by considering the small $Cu$ regime and use \eqref{final_U_eq} to derive an analytical formula for the leading-order swimming speed $U$ of a two-mode squirmer (with $B_1$ and $B_2$ modes)
\begin{equation}\label{lowCu1}
\frac{U}{U_N} = 1+ Cu^2 \left( 1- \beta \right) \frac{\left(n - 1\right)}{2} C_1\left[ 1 + C_2 \alpha^2 \right], 
\end{equation}
where $C_1 = 0.33$ and $C_2 = 2.25$ are numerical constants, and $U_N$ is the Newtonian swimming speed. In a shear-thinning fluid we have $n < 1$ and $\beta < 1$ , and hence we find that this two-mode squirmer can only swim slower than in a Newtonian fluid ($U/U_N <1$) in the small $Cu$ regime. The two-dimensional numerical simulations in \citet{Montenegro_shearthinning} reported that a neutral squirmer ($\alpha = 0$) swims slower in a shear-thinning fluid, which is consistent with our analytical results for a three-dimensional squirmer \eqref{lowCu1} in the small $Cu$ regime, but we also find that the same conclusion of a decreased swimming speed holds for pushers ($\alpha <0$) and pullers ($\alpha > 0$) as well. In contrast to the case of swimming in a viscoelastic fluid, where the pusher and puller attain different velocities given the same magnitude of $\alpha$ \citep{Pullers_vs_Pushers}, \eqref{lowCu1} reveals that in a shear-thinning fluid a pusher and puller have the same swimming velocity because the function for swimming speed is even in $\alpha$; this asymptotic result is verified by numerical simulations to hold for different ranges of $Cu$ and $\beta$ (as shown by the overlapping of the upper and lower triangles in figure \ref{fig1}).

To further characterise the variation of swimming speed, over the full range of $Cu$, we consider the asymptotic limit $\epsilon = 1-\beta\ll 1$, aided by numerical simulations for larger values of $\epsilon$. Biological fluids often have a small viscosity ratio $\beta$ and hence $\varepsilon = 1-\beta$ is typically close to 1. In figure 1a, we present the numerical results for a neutral squirmer, pusher, and puller in the biological limit using the values $\varepsilon = 0.99$ and $n = 0.25$ to emulate human cervical mucus \citep{Rheology_mucus, Rodrigo_lauga}. We demonstrate in the upper inset (neutral squirmer) and lower inset (pusher and puller) in figure 1a that the numerical solutions for the swimming speed ratio converge to the asymptotic solutions (solid line in the upper inset; dashed line in the lower inset) when $\varepsilon \rightarrow 0$. In figure 1b, the results are presented at $\varepsilon = 0.1$ and we note that all qualitative features of the impact of a shear-thinning fluid in the biological limit ($\varepsilon \approx 1$, figure 1a) on the swimming speed are well captured by the asymptotic analysis (when $\varepsilon \ll 1$, figure 1b) and as expected, the numerical simulations (symbols) agree very well with the asymptotic theory (lines) when $\varepsilon$ is small (figure 1b). 

The non-monontonic variation of the swimming speed with $Cu$ may be expected based on the asymptotic behaviour of the constitutive relation discussed at the end of section \ref{sec:CY}. To understand the variation more quantitatively, recall the form of $\mathbf{A}$ in \eqref{beta_expansion} and observe from the integral expression for swimming velocity \eqref{final_U_eq} that at low strain rates, the non-Newtonian contribution $\mathbf{A} \sim \frac{1}{2}Cu^2 (n-1) \vert \dot{\gamma}_0\vert^2 \boldsymbol{\dot{\gamma}}_0$ vanishes as $Cu \rightarrow 0$. At high strain rates, $\mathbf{A} \sim -\boldsymbol{\dot{\gamma}}_0+\left(Cu \vert \dot{\gamma}_0\vert\right)^{n-1} \boldsymbol{\dot{\gamma}}_0$; the first term, $-\boldsymbol{\dot{\gamma}}_0$, vanishes under the integration in \eqref{final_U_eq} \citep{Gwynn_book}, and the remaining term gives a non-Newtonian contribution that vanishes as $Cu\rightarrow \infty$ because for a shear-thinning fluid $n<1$. The swimming speed therefore displays a non-monotonic variation with $Cu$, and since the speed decreases when $Cu$ is small as shown by \eqref{lowCu1}, a minimum swimming speed may be expected to occur at intermediate values of $Cu$ (the `power-law' regime of the model), where the non-Newtonian effect is most significant. However, for a given swimming gait, if the actuation rate of the swimmer is small enough or large enough, the shear thinning fluid may appear to have no effect at all on the swimming speed.

To understand the reduction in swimming speed, inspired by the qualitative descriptions given in \citet{Montenegro_shearthinning}, we look into the thrust and the drag of the swimming problem separately.

\subsection{Drag and thrust}
We separate the swimming problem into a drag problem (a sphere undergoing rigid body translation $\mathbf{U}$ inducing hydrodynamic drag), and a thrust problem (a sphere held fixed undergoing only tangential surface distortions thereby generating thrust). The superposition of these two sub-problems gives the entire swimming problem in a Newtonian fluid in the Stokes regime; this is obviously not the case in a shear-thinning fluid due to its nonlinear constitutive equation. However, by looking at the thrust and the drag problems, one may gain insight into the more complex non-Newtonian swimming problem \citep{Montenegro_shearthinning,qiu2014swimming}. 

We derive the expressions for drag and thrust in a shear-thinning fluid again via the reciprocal theorem approach (section \ref{reciprocal_theorem}) by utilizing the solution to the resistance problem in a Newtonian fluid. The drag force on a sphere moving with a velocity $\textbf{U}$ in weakly shear-thinning fluid is given by $\textbf{F}_{D} = - 6 \pi \textbf{U} - \frac{3}{4} \varepsilon \int_V \mathbf{A}_D : \left( 1+ \frac{1}{6} \nabla^2 \right)\nabla {\mathbf{G}}$, where $\mathbf{A}_D$ is formed by the solution to the Newtonian drag problem. Similarly, the thrust force generated by a sphere held stationary with surface actuation $\textbf{u}^S$ in a weakly shear-thinning fluid is given by $\textbf{F}_{T} = -\frac{3}{2} \int_S \textbf{u}^S  dS - \frac{3}{4} \varepsilon \int_V \mathbf{A}_T : \left( 1+ \frac{1}{6} \nabla^2 \right)\nabla {\mathbf{G}}$, where $\mathbf{A}_T$ is formed by the solution to the Newtonian thrust problem.

\begin{figure}
\centering
\begin{minipage}{.49\textwidth}
  \centering
  \includegraphics[scale = 0.25, trim = 0 30 50 70, clip ]{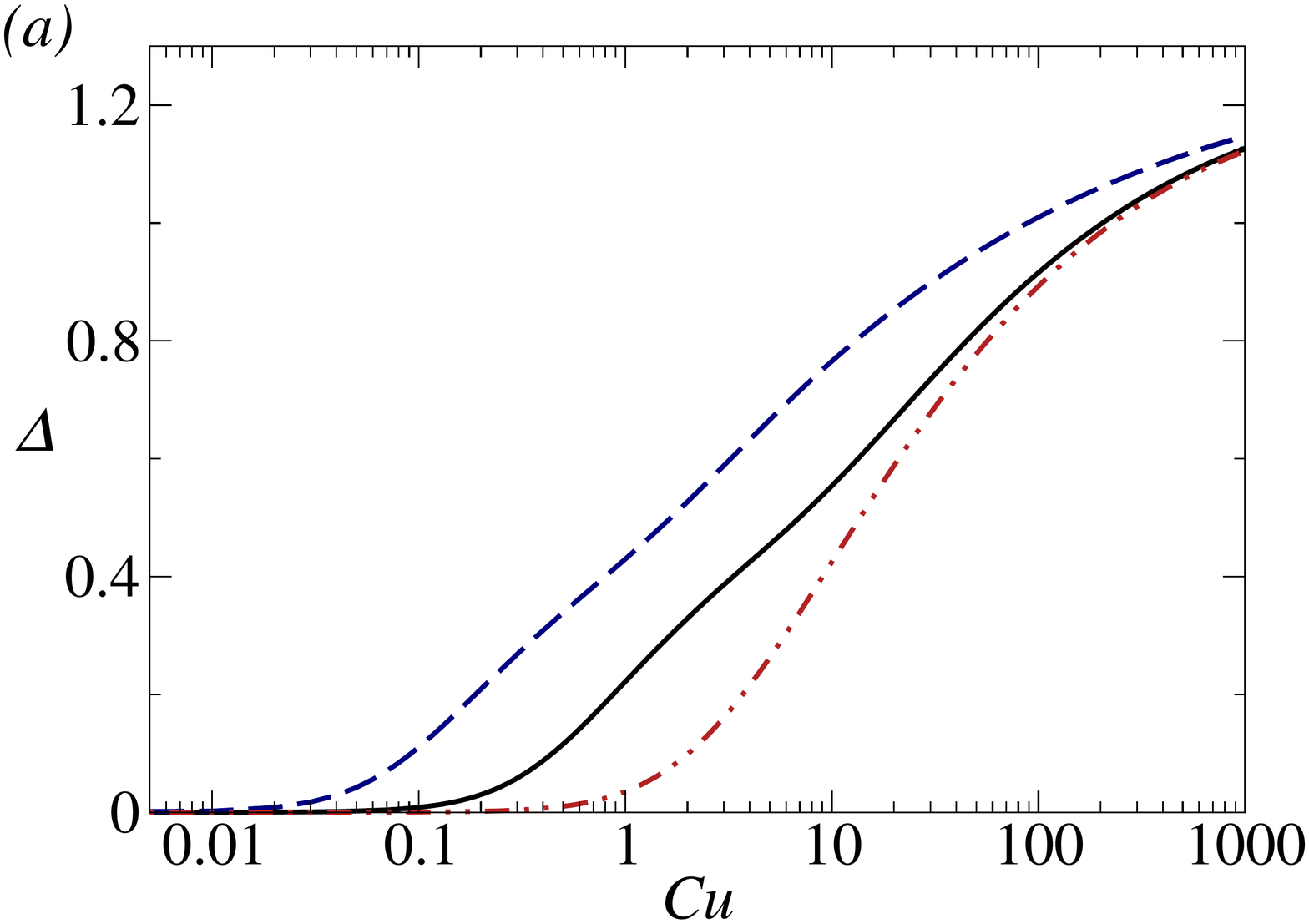}
  \end{minipage}%
\begin{minipage}{.4\textwidth}
  \centering
  \includegraphics[scale= 0.25, trim = 0 30 50 70, clip]{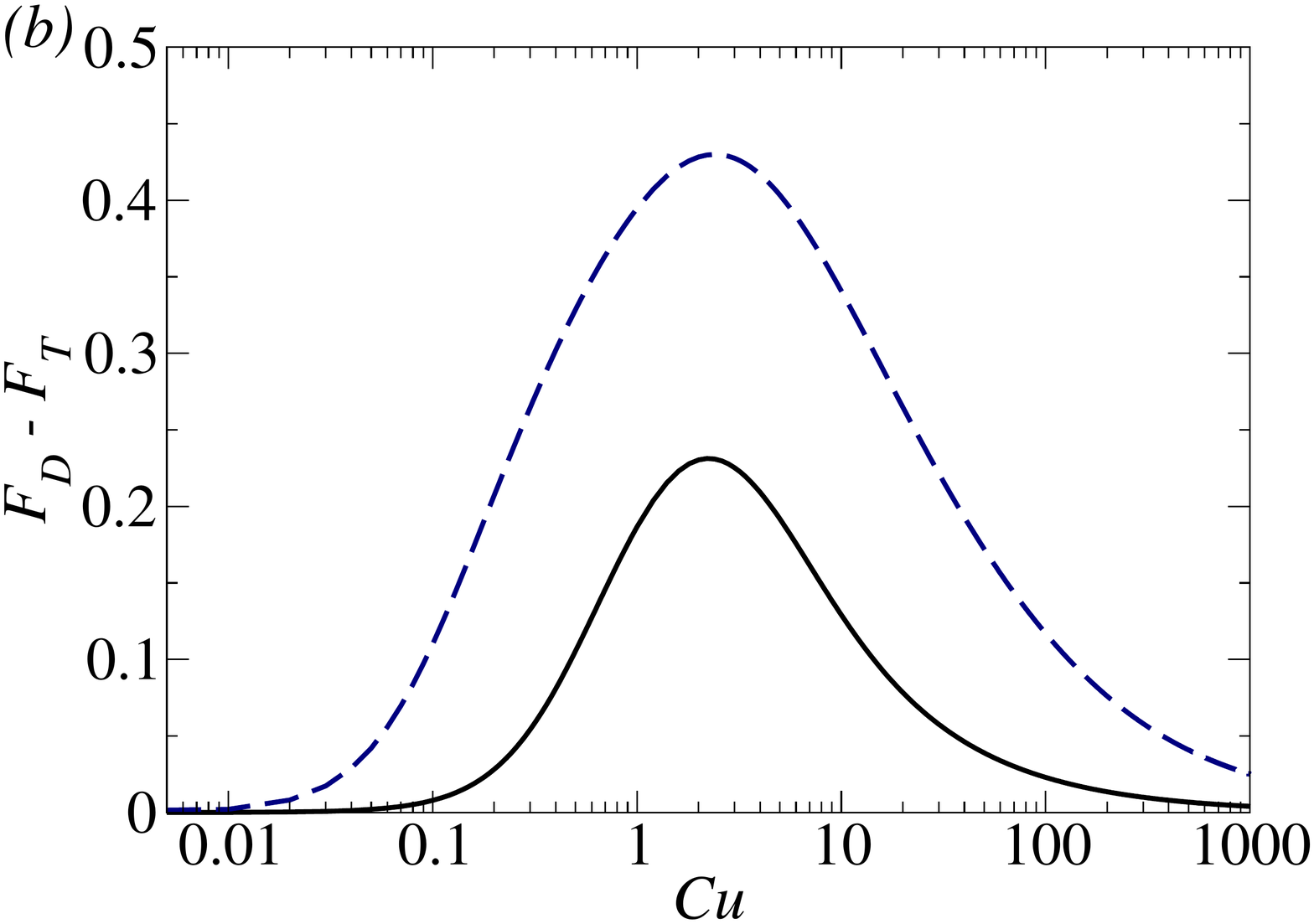}
 \end{minipage}
\caption{In \textit{(a)}, the symbol $ \Delta$ denotes the difference between drag, or thrust, in a shear-thinning fluid and a Newtonian fluid: 
 the dash-dot curve represents the drag reduction in a shear-thinning fluid compared with a Newtonian fluid, while the solid and dashed curves represent loss in thrust for squirmers with $\alpha=0$ and $\alpha = \pm 5$ respectively ($n=0.25$, $\varepsilon = 0.1$). In \textit{(b)} we show that the difference between drag and thrust is positive both for a neutral squirmer (solid) and a pusher/puller (dashed) in a shear-thinning fluid. All quantities are dimensionless}.
  \label{fig2}
\end{figure}

One could expect a drag reduction when a rigid sphere is pulled with a constant velocity through a shear-thinning fluid since the fluid viscosity is reduced by the fluid straining motion. However, it is interesting to see in figure \ref{fig2}a that the thrust reduction caused by the shear-thinning rheology is larger than the drag reduction for a large range of $Cu$.  This more severe reduction in thrust than drag then suggests slower swimming speeds compared with the Newtonian case, which correctly predicts the trend found by detailed calculations (figure \ref{fig1}). In addition, for very small or large values of $Cu$, the difference between the magnitudes of drag and thrust ($F_D - F_T$) vanishes  as shown in figure \ref{fig2}b, respecting the limits where the swimming speed should recover the Newtonian value (figure \ref{fig1}).

Although conceptually intuitive, the drag and thrust decomposition is not complete as it neglects the contribution of non-linear products in the non-Newtonian stress to the full swimming problem, namely $\mathbf{A} \ne \mathbf{A}_T+\mathbf{A}_D$ owing to the non-linearity in the constitutive equation. We will give a counter-example in section \ref{sec:modes} below to illustrate a scenario when these intuitive arguments fail.

\subsection{Addition of other squirming modes}\label{sec:modes}

The results from the detailed asymptotic and numerical analysis as well as the intuitive model for a two-mode squirmer seem to suggest that the shear-thinning rheology acts to hinder the locomotion. This raises the simple question of whether this conclusion still holds if other modes of surface actuation are present. The picture is clear for a Newtonian fluid: only the $B_1$ mode contributes to swimming and adding other modes does not alter the swimming speed. However, can the shear-thinning rheology render other modes, typically not considered in the Newtonian analysis, effective for propulsion? We address these questions using the asymptotic and numerical tools developed in the previous sections.

\begin{figure}
\centering
\begin{minipage}{.49\textwidth}
  \centering
  \includegraphics[scale = 0.25, trim = 0 30 50 70, clip ]{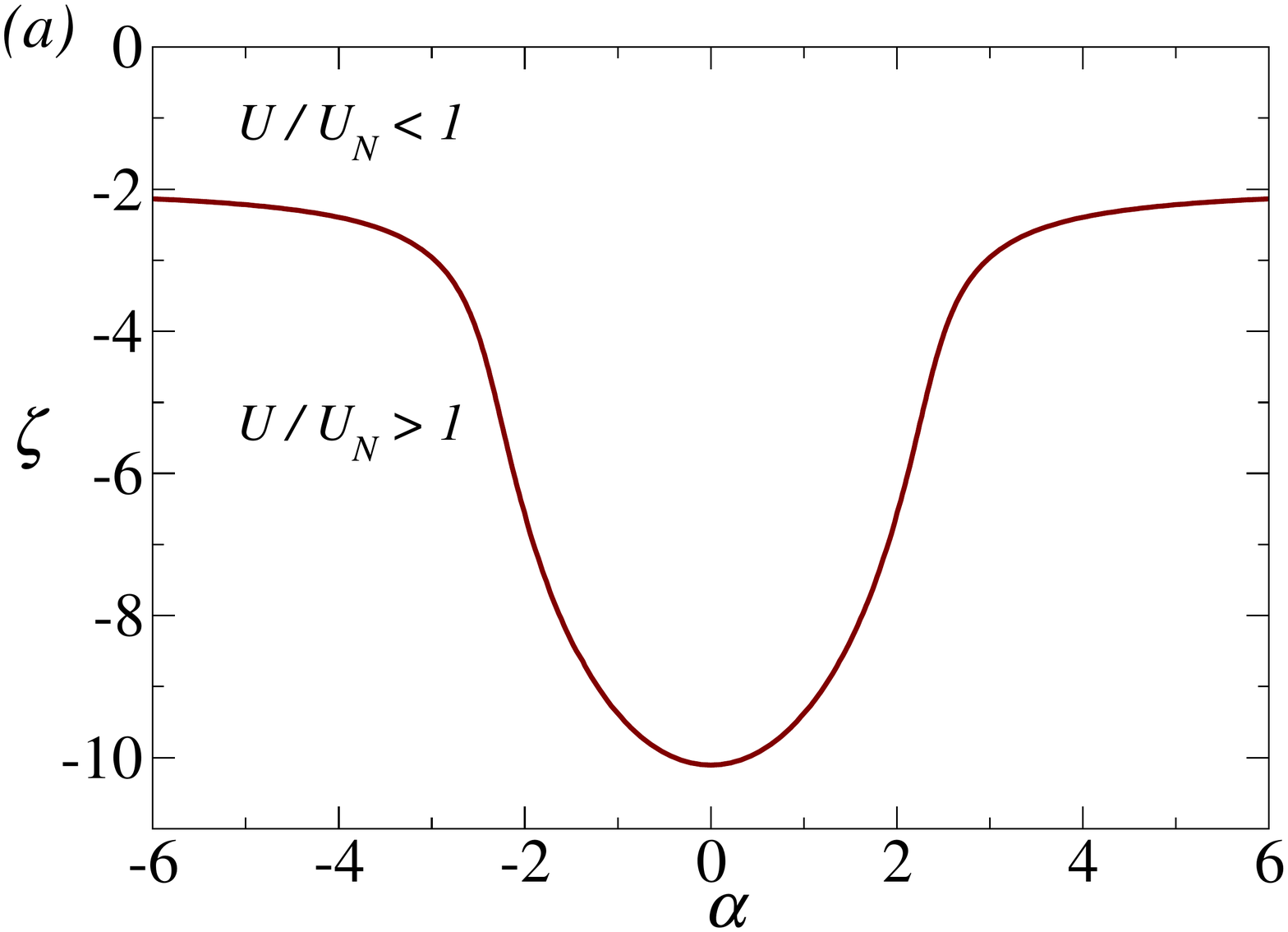}
  \end{minipage}%
\begin{minipage}{.4\textwidth}
  \centering
  \includegraphics[scale= 0.25, trim = 0 30 50 70, clip]{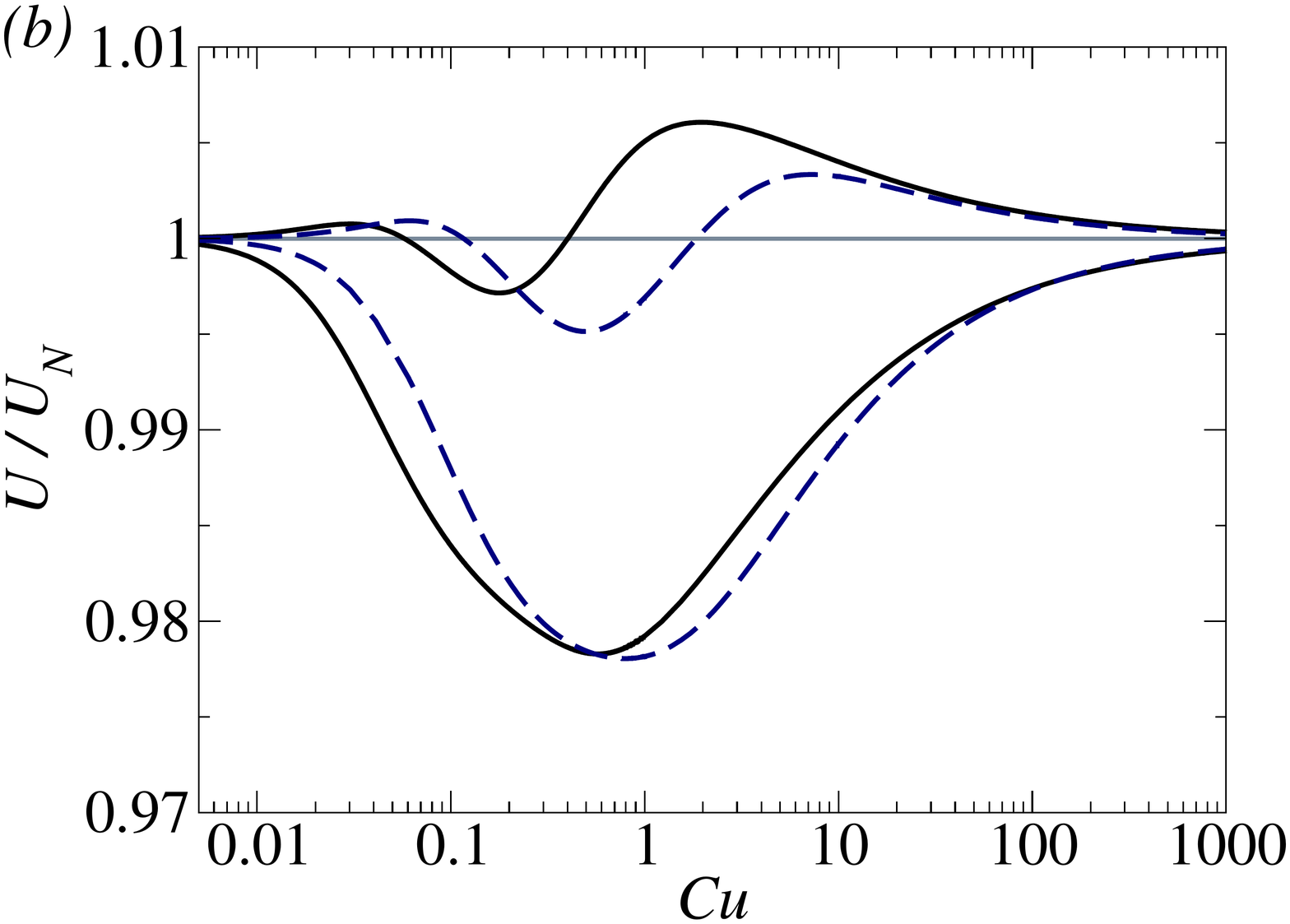}
 \end{minipage}
 \caption{In \textit{(a)} we show the level set curve below which faster swimming occurs for specific values of $\alpha$ and $\zeta =B_3/B_1$ when $Cu\ll 1$. In \textit{(b)} we show the swimming speed of a neutral squirmer (solid line) and a pusher/puller (dashed line) at values of $\zeta$ chosen below the curve in \textit{(a)} so that the swimming speed is larger than the Newtonian value at small $Cu$ (the upper solid line: $\alpha=0$, $\zeta=-15$ ; the upper dashed line: $\alpha=\pm 5$, $\zeta=-4$ ); conversely, with values of $\zeta$ above the curve in \textit{(a)} the swimmers always swim slower than in a Newtonian fluid as shown (lower solid line: $\alpha=0$, $\zeta=15$; lower dashed lines: $\alpha=\pm 5$, $\zeta=4$). Here $\varepsilon = 0.1$, $n = 0.25$.}
  \label{fig3}
\end{figure}

We first note that the $B_3$ mode alone leads to locomotion in a shear-thinning fluid in stark contrast to a Newtonian fluid as only the $B_1$ mode has a non-zero surface average (see \eqref{final_U_eq}). Indeed any odd mode alone may lead to locomotion in a shear-thinning fluid (even modes alone do not swim by symmetry). We also find quite distinctive behaviour when the $B_3$ mode is combined with other modes. In the $Cu \ll 1$ regime, we can derive an analytical expression allowing us to predict the values of $\alpha$ and $\zeta = B_3 / B_1$ for faster or slower swimming. To quadratic order in $Cu$, we find 
\begin{align}
\frac{U}{U_N} = 1+ Cu^2 \left(1-\beta\right) \frac{\left(n-1\right)}{2} C_1\left[ 1+ C_2\left( 1 +  C_3 \zeta \right) \alpha^2+ C_4\left(C_5\zeta^2+C_6\zeta   -1\right)\zeta \right], 
\label{3mode_carreau}
 \end{align}
where the additional numerical constants are given by $C_3 = 0.51$, $C_4 = 0.70$, $C_5 = 0.18$, and $C_6 = 1.66$. Again the swimming speed is even in $\alpha$ and we recover \eqref{lowCu1} when $\zeta=0$ as expected. From \eqref{3mode_carreau} we can predict that faster swimming ($U/U_N >1$) will occur if 
\begin{align}
\left.\frac{\partial ^2 U}{\partial Cu^2}(\alpha,\zeta)\right|_{Cu=0}>0,
\end{align}
in other words when the term in the square brackets in \eqref{3mode_carreau} is negative. In figure \ref{fig3}a we plot the level set curve below which faster swimming occurs in the small $Cu$ regime; we find that this can only occur when $\zeta$ is negative for any $\alpha$. For example, when $\alpha = 0$ we must have $\zeta < -10.11$, while $\alpha = \pm 5$, $\zeta < -2.22$ leads to faster swimming.

In figure \ref{fig3}b we show the variation of swimming speed for two swimmers with $\alpha$ and $\zeta$ chosen below the level set curve (the upper solid and dashed lines) and two swimmers with $\alpha$ and $\zeta$ chosen above the level set curve (the lower solid and dashed lines). We note that the swimming speeds of the faster swimmers in the small $Cu$ regime experience a subsequent fall below the Newtonian value and then a rise above it as $Cu$ increases, before asymptoting to the Newtonian swimming speed at high $Cu$. This indicates that microorganisms (with a given swimming gait) can swim both faster, and slower, than in a Newtonian fluid depending on the actuation rate of that gait. In contrast, the two swimmers that swim slower in a Newtonian fluid in the small $Cu$ regime remain slower for larger $Cu$ with a non-monotonic variation similar to that observed previously (see figure \ref{fig1}). These results also hold qualitatively for large values of $\epsilon$.

We emphasize that the thrust and drag reduction model is unable to explain the faster swimming speed with the addition of a $B_3$ mode because in these cases the thrust still decreases more than the drag over a wide range of Carreau numbers, if they are considered separately. This serves as a counter-example demonstrating how analyzing drag and thrust separately may not adequately describe swimming in complex media. This fact points specifically to the interaction between thrust and drag fields, due to the nonlinearity in the constitutive equation, as the cause of faster than Newtonian swimming.

\section{Conclusion}
We show in this paper that shear-thinning rheology affects a squirmer with a prescribed swimming gait in nontrivial and surprising ways; we predict, analytically, instances of both faster and slower swimming than in a Newtonian fluid depending on the details of the prescribed boundary conditions. Indeed we demonstrate that even with the same squirming modes a squirmer can swim faster or slower depending on its rate of actuation. In general, these results point to the importance of both the spatial and temporal details of the swimming gait of a microorganism and ultimately the difficulty in predicting the resulting effect of the non-Newtonian fluid \textit{a priori}. In light of this, an important next step would be to incorporate models of internal force generation for biological swimmers and determine how fluid rheology affects the resultant gait itself in concert with propulsion. Finally, we remark that the drag and thrust decomposition of the swimming problem is indeed effective in Newtonian fluids and may also be insightful in complex fluids in some instances, but one should use caution when extending the results to non-Newtonian swimming as the inherent non-linearity of the problem can be significant enough for a Newtonian-like decomposition to yield qualitatively flawed predictions as illustrated by the example we provide.

\section*{Acknowledgements}
G.J.E. acknowledges funding from NSERC grant RGPIN-2014-06577. 
L.Z. acknowledges the financial support from ERC grant SIMCOMICS-280117.
G.J.E. and C.D. gratefully acknowledge support from Professor G.M. Homsy. 

\newpage

\bibliography{squirmer}

\end{document}